\documentclass[twocolumn,prb]{revtex4}
\usepackage[T1]{fontenc}
\usepackage{amsmath,amsbsy,amssymb,graphicx}
\usepackage{times}

\let\mathbf=\boldsymbol
\begin{document}

\title{{\Large Spin-Valley Optical Selection Rule and Strong Circular Dichroism in Silicene}}
\author{Motohiko Ezawa}
\affiliation{Department of Applied Physics, University of Tokyo, Hongo 7-3-1, 113-8656,
Japan }

\begin{abstract}
Silicene (a monolayer of silicon atoms) is a topological insulator, which
undergoes a topological phase transition to a band insulator under an
external electric field. The spin polarization is unique and opposite at the
K and K' points due to the spin-orbit coupling. Accordingly, silicene
exhibits a strong circular dichroism with respect to optical absorption,
obeying a certain spin-valley selection rule. It is remarkable that this
selection rule is drastically different between these two types of
insulators owing to a band inversion taking place at the phase transition
point. Hence we can tell experimentally whether silicene is in the
topological or band insulator phase by circular dichroism. Furthermore the
selection rule enables us to excite electrons with definite spin and valley
indices by optical absorption. Photo-induced current is spin polarized,
where the spin direction is different between the topological and band
insulators. It is useful for future spintronics applications.
\end{abstract}

\maketitle


\address{{\normalsize Department of Applied Physics, University of Tokyo, Hongo
7-3-1, 113-8656, Japan }}

\textbf{Introduction:} Silicene consists of a honeycomb lattice of silicon
atoms with buckled sublattices made of A sites and B sites. The states near the
Fermi energy are $\pi $ orbitals residing near the K and K' points at
opposite corners of the hexagonal Brillouin zone. Silicene has recently been
synthesized\cite{GLayPRL,Kawai,Takamura} and attracted much attention\cite%
{Ciraci,LiuPRL,EzawaNJP,EzawaAQHE}. The low-energy dynamics in the K and K'
valleys is described by the Dirac theory as in graphene. However, Dirac
electrons are massive due to a relatively large spin-orbit (SO) gap of $1.55$%
meV in silicene. It is remarkable that the mass can be controlled\cite%
{EzawaNJP} by applying the electric field $E_{z}$ perpendicular to the
silicene sheet.

A novel feature is that silicene is a topological insulator\cite{LiuPRL},
which is characterized by a full insulating gap in the bulk and helical
gapless edges\cite{Hasan,Qi}. It undergoes a topological phase transition
from a topological insulator to a band insulator as $|E_{z}|$ increases and
crosses the critical field $E_{\text{cr}}$, as has been shown by examining
numerically the emergence of the helical zero energy modes in silicene
nanoribbons\cite{EzawaNJP} and also by calculating the topological numbers%
\cite{EzawaDiamag}. It is an intriguing problem how to detect experimentally
whether an insulator is topological or not just by the bulk property.

\begin{figure}[t]
\centerline{\includegraphics[width=0.45\textwidth]{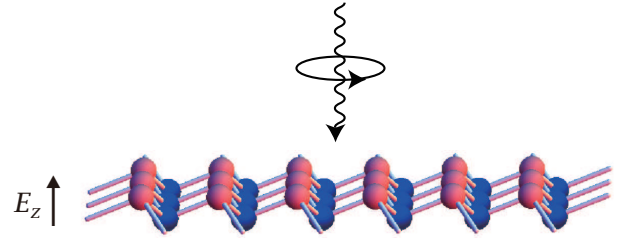}}
\caption{(Color online) Illustration of the buckled honycomb lattice of
silicene. A honeycomb lattice is distorted due to a large ionic radius of a
silicon atom and forms a buckled structure. The A and B sites form two
sublattices separated by a perpendicular distance $2\ell $. The structure
generates a staggered sublattice potential in the electric field $E_{z}$.
Silicene undergoes a phase transition from a topological insulator to a band
insulator. It is possible to discreminate between these two types of
insulators by circular dichroism.}
\label{FigBuckl}
\end{figure}

The interaction of light with matter depends on the polarization of the
photons in general. When the response of a system to the left- and
right-handed circularly polarized light is different, the phenomenon is
referred to as circular dichroism. The circular dichroism has been shown to
be essential to analyze the valley-dependent interplay of electrons in
graphene\cite{Xiao07,Yao08} as well as in monolayer dichalcogenides\cite%
{Xiao,Cao}.

In this paper, we investigate optical absorptions in silicene. The crucial
property is the spin polarization is unique and opposite at the K and K'
points due to the SO coupling for $E_{z}\neq 0$. Hence silicene exhibits a
strong circular dichroism, where the optical absorption of circularly
polarized light strongly depends on the spin and valley. Furthermore, the
optical selection rules are entirely different between a topological
insulator and a band insulator, as enables us to tell whether silicene is in
the topological or band insulator phase. The difference originates in a band
inversion taking place at the phase transition point. It is remarkable that
we can detect a topological phase transition by the change of circular
dichroism as $|E_{z}|$ crosses the critical field $E_{\text{cr}}$. It is
also remarkable that, by irradiating right-handed circularly polarized
light, for instance, we can selectively excite up spin at the K point. By
applying an in-plane electric field, photo-exicited spin-polarized charges
can be extracted. We can determine whether the system is topological or band
insulator by detecting the spin direction of the photo-induced current.

\textbf{Low-Energy Dirac Theory:} We take a silicene sheet on the $xy$%
-plane, and apply the electric field $E_{z}$ perpendicular to the plane. Due
to the buckled structure the two sublattice planes are separated by a
distance, which we denote by $2\ell $ with $\ell =0.23$\AA\ , as illustrated
in Fig.\ref{FigBuckl}. It generates a staggered sublattice potential $%
\varpropto 2\ell E_{z}$ between silicon atoms at A sites and B sites.

We analyze the physics of electrons near the Fermi energy, which is
described by Dirac electrons near the K and K' points. We also call them the
K$_{\xi }$ points with $\xi =\pm $. The effective Dirac Hamiltonian in the
momentum space reads\cite{EzawaAQHE}%
\begin{align}
H_{\xi }=& \hbar v_{\text{F}}\left( \xi k_{x}\tau _{x}+k_{y}\tau _{y}\right)
+\lambda _{\text{SO}}\sigma _{z}\xi \tau _{z}-\ell E_{z}\tau _{z}  \notag \\
& +a\xi \tau _{z}\lambda _{\text{R2}}\left( k_{y}\sigma _{x}-k_{x}\sigma
_{y}\right)  \notag \\
& +\lambda _{\text{R1}}\left( E_{z}\right) (\xi \tau _{x}\sigma _{y}-\tau
_{y}\sigma _{x})/2,  \label{DiracHamilSilic}
\end{align}%
where $\sigma _{a}$ and $\tau _{a}$ are the Pauli matrices of the spin and
the sublattice pseudospin, respectively. We explain each term. The first
term arises from the nearest-neighbor hopping, where $v_{\text{F}}=\frac{%
\sqrt{3}}{2}at=5.5\times 10^{5}$m/s is the Fermi velocity with the transfer
energy $t=1.6$eV and the lattice constant $a=3.86$\AA . The second term
represents the effective SO coupling\cite{KaneMele,LiuPRB} with $\lambda _{%
\text{SO}}=3.9$meV. The third term represents the staggered sublattice
potential\cite{EzawaNJP} in electric field $E_{z}$. The forth term
represents the second Rashba SO coupling with $\lambda _{\text{R2}}=0.7$meV
associated with the next-nearest neighbor hopping term\cite{LiuPRB}. The
fifth term represents the first Rashba SO coupling associated with the
nearest neighbor hopping, which is induced by external electric field\cite%
{Hongki,Tse}. It satisfies $\lambda _{\text{R1}}(0)=0$ and becomes of the
order of $10\mu $eV at the electric field $E_{\text{c}}=\lambda _{\text{SO}%
}/\ell =17$meV\AA $^{-1}$. Its effect is negligible as far as we have
checked. Although we include all terms in numerical calculations, in order
to simplify the formulas and to make the physical picture clear, we set $%
\lambda _{\text{R1}}(E_{z})=0$ in all analytic formulas.

\begin{figure}[t]
\centerline{\includegraphics[width=0.5\textwidth]{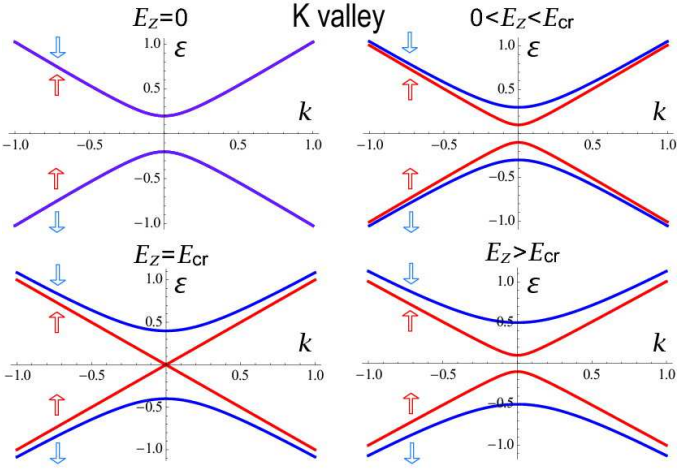}}
\caption{(Color online) Band structure of silicene in the K valley under
electric field $E_{z}$. The spins of electrons on the red (blue) band are
almost up (down) polirized. They are exactly polorized if $\protect\lambda _{%
\text{R2}}=0$. Although the energy spectrum looks similar between the
topological and band insulators, the pseudospin $t_{z}$ is opposite at the K
point. Indeed, there is a band inversion at the K point: See Fig.\protect\ref%
{FigOptLevel}. The spin direction is opposite in the K' valley. The units of
the vertical and holizontal axes are arbitrary.}
\label{FigDiracBand}
\end{figure}

There are four bands in the energy spectrum (Fig.\ref{FigDiracBand}). The
band gap is located at the K and K' points. At these points the energy is
given by%
\begin{equation}
E(s_{z},t_{z})=\lambda _{\text{SO}}s_{z}t_{z}-\ell E_{z}t_{z},
\label{DiagoE}
\end{equation}%
with the spin $s_{z}=\pm 1$ and the sublattice pseudospin $t_{z}=\pm 1$.
They are good quantum numbers at the K and K' points. The spin $s_{z}$ is an
almost good quantum number even away from the K and K' points because $%
\lambda _{\text{R2}}$ is a small quantity. On the other hand, the pseudospin 
$t_{z}$ is strongly broken away from the K and K' points for $E_{z}\neq 0$.

The gap is given by $2|\Delta _{s}\left( E_{z}\right) |$ with%
\begin{equation}
\Delta _{s}\left( E_{z}\right) =-\lambda _{\text{SO}}+s\ell E_{z},
\label{gapDiracX}
\end{equation}%
where $s=\pm 1$ is the spin-chirality. It is given by $s=\xi s_{z}$ when the
spin $s_{z}$ is a good quantum number. As $|E_{z}|$ increases, the gap
decreases linearly, and closes at the critical point $|E_{z}|=E_{\text{cr}}$
with%
\begin{equation}
E_{\text{cr}}=\pm \lambda _{\text{SO}}/\ell =\pm 17\text{meV/\AA },
\label{StepA}
\end{equation}%
and then increases linearly. Silicene is a semimetal due to gapless modes at 
$|E_{z}|=E_{\text{cr}}$, while it is an insulator for $|E_{z}|\neq E_{\text{%
cr}}$.

\begin{figure}[t]
\centerline{\includegraphics[width=0.3\textwidth]{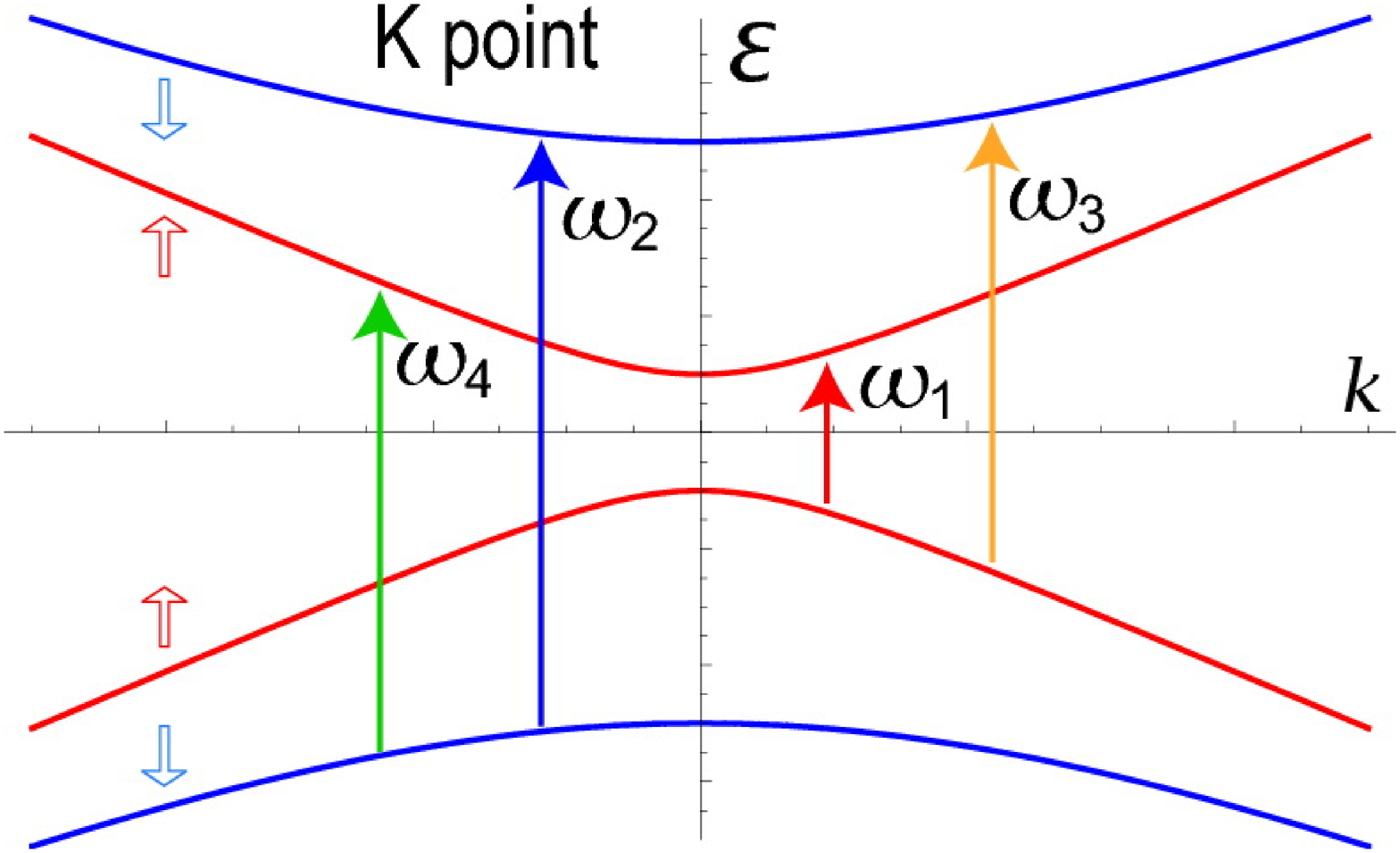}}
\caption{(Color online) Illustration of \ photo-induced transition from the
valence band to the conduction band at the K point. There are four ways of
transitions, which we label as $\protect\omega _{i}$. The arrows $\protect%
\omega _{1}$ and $\protect\omega _{2}$ indicate the optical transition of
high intensity proportional to $v_{\text{F}}^{2}$ and the arrows $\protect%
\omega _{3}$ and $\protect\omega _{4}$ indicate the optical transition of
low intensity proportional to $\protect\lambda _{\text{R2}}^{2}$. We call
the transition $\protect\omega _{1}$ as the fundamental transition.}
\label{FigTransition}
\end{figure}

\textbf{Optical properties:} We explore optical interband transitions from
the state $\left\vert u_{\text{v}}\left( k\right) \right\rangle $ in the
valence band to the state $\left\vert u_{\text{c}}\left( k\right)
\right\rangle $ in the conduction band. There are four transitions, which we
label as $\omega _{1}$, $\omega _{2}$, $\omega _{3}$ and $\omega _{4}$, as
depicted in Fig.\ref{FigTransition}. We call $\omega _{1}$ as the
fundamental transition.

We consider a beam of circularly polarized light irradiated onto the
silicene sheet. The corresponding electromagnetic potential is given by 
\begin{equation}
\mathbf{A}(t)=(A\sin \omega t,A\cos \omega t).
\end{equation}%
The electromagnetic potential is introduced into the Hamiltonian (\ref%
{DiracHamilSilic}) by way of the minimal substitution, that is, replacing
the momentum $\hbar k_{i}$ with the covariant momentum $P_{i}\equiv \hbar
k_{i}+eA_{i}$. The resultant Hamiltonian simply reads%
\begin{equation}
H_{\xi }\left( A\right) =H_{\xi }+\mathcal{P}_{x}^{\xi }A_{x}+\mathcal{P}%
_{y}^{\xi }A_{y},
\end{equation}%
with (\ref{DiracHamilSilic}) and%
\begin{align}
\mathcal{P}_{x}^{\xi }& =\frac{1}{\hbar }\frac{\partial H_{\xi }}{\partial
k_{x}}=v_{\text{F}}\xi \tau _{x}-\frac{a\lambda _{\text{R2}}}{\hbar }\xi
\tau _{z}\sigma _{y},  \notag \\
\mathcal{P}_{y}^{\xi }& =\frac{1}{\hbar }\frac{\partial H_{\xi }}{\partial
k_{y}}=v_{\text{F}}\tau _{y}+\frac{a\lambda _{\text{R2}}}{\hbar }\xi \tau
_{z}\sigma _{x},
\end{align}%
since the Dirac Hamiltonian is linear in the momentum. It is notable that
the formula does not contain the SO coupling $\lambda _{\text{SO}}$. We
conclude from this formula that the kinetic term ($\varpropto v_{\text{F}}$)
induces an interband transition between electrons carrying the same spin
while the Rashba term ($\varpropto \lambda _{\text{R2}}$) induces an
interband transition between electrons carrying the opposite spins.

\begin{figure}[t]
\centerline{\includegraphics[width=0.5\textwidth]{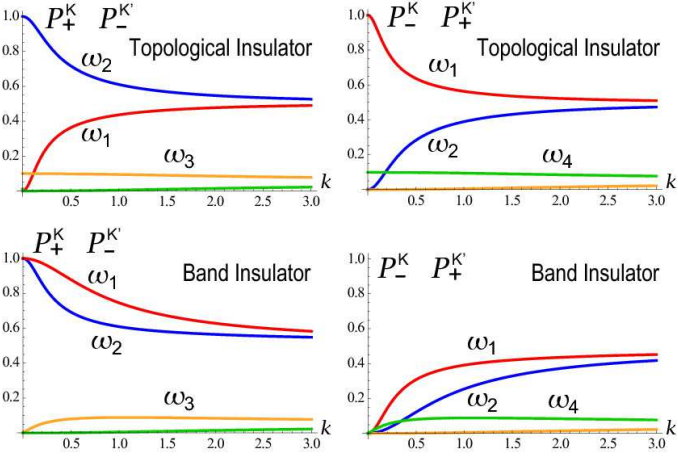}}
\caption{(Color online) The $k$-resolved optical absorption $\left\vert
P_{\pm }^{\protect\kappa }\left( k\right) \right\vert /m_{0}v_{\text{F}}$
near the K and K' points. The horizontal axis is momentum $k$. We have
chosen $E_{z}=E_{\text{cr}}/2$ for a topological insulator and $E_{z}=2E_{%
\text{cr}}$ for a band insulator. Notice the behavior of the interband
transition $\protect\omega _{1}$, which changes drastically in the
topological and band insulators. For illustration we have taken $a\protect%
\lambda _{\text{R2}}/\hbar =0.1v_{\text{F}}$, though the actual value is
about $10^{-3}v_{\text{F}}$.}
\label{FigOptK}
\end{figure}

The coupling strength with optical fields of the\ right($+$) or left($-$)
circular polarization is given by $\mathcal{P}_{\pm }^{\xi }\left( k\right) =%
\mathcal{P}_{x}^{\xi }\left( k\right) \pm i\mathcal{P}_{y}^{\xi }\left(
k\right) $. They are written in terms of the ladder operator of spins and
pseudospins.%
\begin{eqnarray}
\mathcal{P}_{\pm }^{K} &=&v_{\text{F}}\tau _{\pm }\pm \frac{ia\lambda _{%
\text{R2}}}{\hbar }\tau _{z}s_{\pm }, \\
\mathcal{P}_{\pm }^{K^{\prime }} &=&-v_{\text{F}}\tau _{\mp }\mp \frac{%
ia\lambda _{\text{R2}}}{\hbar }\tau _{z}s_{\pm },
\end{eqnarray}%
where $s_{\pm }=s_{x}\pm is_{y}$ and $\tau _{\pm }=\tau _{x}\pm i\tau _{y}$.

The matrix element between the initial state and the final state in the
photoemission process is given by 
\begin{equation}
P_{\pm }^{\xi }\left( k\right) \equiv m_{0}\left\langle u_{\text{c}}\left(
k\right) \right\vert \frac{1}{\hbar }\frac{\partial H_{\xi }}{\partial
k_{\pm }}\left\vert u_{\text{v}}\left( k\right) \right\rangle ,
\label{OpticAbsor}
\end{equation}%
where $m_{0}$ is the free electron mass. Here, $|P_{\pm }^{\eta }\left(
k\right) |$ is called the optical absorption. There exist the relations,%
\begin{equation}
P_{+}^{K}\left( k\right) =P_{-}^{K^{\prime }}\left( k\right) ,\qquad
P_{-}^{K}\left( k\right) =P_{+}^{K^{\prime }}\left( k\right) ,
\end{equation}%
reflecting the time-reversal symmetry. The right-handed circular
polarization at the K point and the left-handed circular polarization at the
K' point are equal.

The wave functions $|u_{\text{v}}\left( k\right) \rangle $ and $|u_{\text{c}%
}\left( k\right) \rangle $ are obtained explicitly by diagonalizing the
Hamiltonian (\ref{DiracHamilSilic}). We have calculated numerically the
optical absorption $|P_{\pm }^{\xi }\left( k\right) |$ as a function of $k$
near the K and K' points, which we display in Fig.\ref{FigOptK}.

Let us investigate the optical absorption at $k=0$ in detail, since we can
obtain a clear physical picture at the K and K' points with the aid of
analytical formulas. The Hamiltonian (\ref{DiracHamilSilic}) is diagonal at
the K and K' points with the eigenvalues given by (\ref{DiagoE}) and the
eigenfunctions given by $(1,0,0,0)^{t}$ and so on. For $E_{z}>0$, the order
of the energy level is%
\begin{equation}
E(-1,-1)>E(1,1)>E(1,-1)>E(-1,1)
\end{equation}%
for the topological insulator, and%
\begin{equation}
E(-1,-1)>E(1,-1)>E(1,1)>E(-1,1)
\end{equation}%
for the band insulator. It is to be emphasized that the two bands near the
Fermi level are inverted between the topological and band insulators, since
the pseudospin $t_{z}$ is flipped as $E_{z}$ exceeds the critical point $E_{%
\text{cr}}$.

\begin{figure}[t]
\centerline{\includegraphics[width=0.5\textwidth]{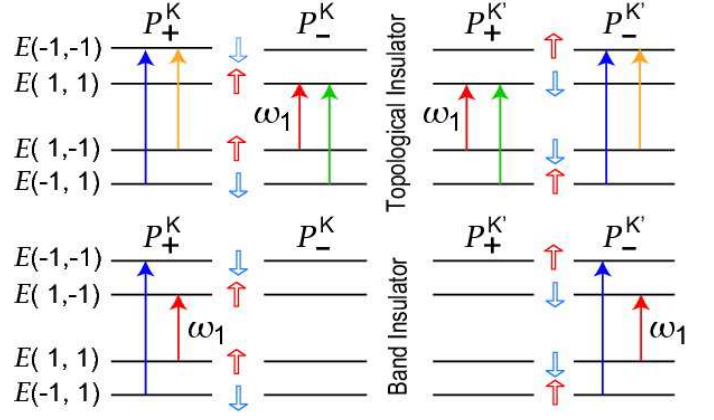}}
\caption{(Color online) Spin-valley optical selection rules at the K and K'
points ($k=0$). The band is indexed by $E(s_{z},t_{z})$. The two bands $%
E(1,1)$ and $E(1,-1)$ are inverted between the topological and band
insulators, as leads to a different circular dichroism to them.}
\label{FigOptLevel}
\end{figure}

The band inversion leads to a different circular dichroism. This is because
the operator $\mathcal{P}_{\pm }$ is not Hermitian: $\mathcal{P}_{\pm }$
describes an optical absorption process, while $\mathcal{P}_{\pm }^{\dagger
} $ describes an optical emission process. Furthermore, an optical
absorption occurs only when the energy of the initial state is lower than
that of the final state. Thus the optical absorption obeys a strong
spin-valley coupled selection rule (Fig.\ref{FigOptLevel}). In conclusion,
the fundamental optical absorption $\omega _{1}$ (indicated by red arrow) is
different whether the system is a topological insulator or a band insulator.
For example, the right-circular polarized light at the K point is absorbed
only when the system is a band insulator, while the left-circular polarized
light at the K point is absorbed only when the system is a topological
insulator. Furthermore the spin-flipped interband transitions occur only
when the system is a topological insulator.

\begin{figure}[t]
\centerline{\includegraphics[width=0.5\textwidth]{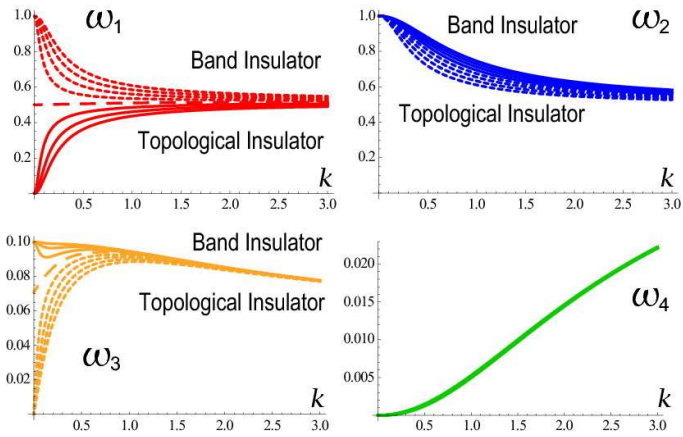}}
\caption{(Color online) The $k$-resolved optical absorption $\left\vert
P_{+}^{K}\left( k\right) \right\vert $ for $E_{z}/E_{\text{cr}%
}=1/4,1/2,3/4,1,5/4,3/2,7/4,2$. The solid (dotted) lines are for topological
(band) insulators, while the dashed line is for the phase transition point ($%
E_{z}=E_{\text{cr}}$).}
\label{FigOptE}
\end{figure}

We show the optical absorption for various electric field in Fig.\ref%
{FigOptE}. If we neglect the Rashba terms ($\lambda _{\text{R2}}=0$) we are
able to obtain an analytic formula for the transitions $\omega _{1}$ and $%
\omega _{2}$ near the K$_{\xi }$ point, 
\begin{equation}
\left\vert P_{\pm }^{\xi }\left( k\right) \right\vert ^{2}=m_{0}^{2}v_{\text{%
F}}^{2}\left( 1\pm \xi \frac{\Delta _{s}\left( E\right) }{\sqrt{\Delta
_{s}^{2}\left( E\right) +4a^{2}t^{2}k^{2}}}\right) ^{2},
\end{equation}%
where $s=+1$ for $\omega _{1}$ and $s=-1$ for $\omega _{2}$. When the
electric field is critical, $E_{z}=E_{\text{cr}}$, where $\Delta _{s}=0$,
the optical absorption of the fundamental transition $\omega _{1}$ becomes a
constant, 
\begin{equation}
\left\vert P_{\pm }^{\xi }\left( k\right) \right\vert ^{2}=m_{0}^{2}v_{\text{%
F}}^{2},
\end{equation}%
which is common to the K and K' valleys. This property holds as it is even
for $\lambda _{\text{R2}}\neq 0$, as seen in Fig.\ref{FigOptE}.%

\begin{figure}[t]
\centerline{\includegraphics[width=0.5\textwidth]{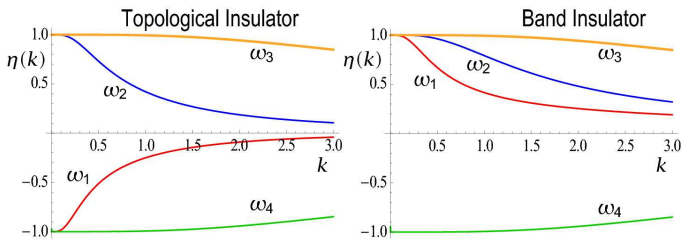}}
\caption{(Color online) The $k$-resolved circular polarization $\protect\eta %
\left( k\right) $ for the interband transitions $\protect\omega _{i}$ in
topological and band insulators.}
\label{FigPorali}
\end{figure}

The $k$-resolved optical polarization $\eta (k)$ is given by%
\begin{equation}
\eta ^{\xi }(k)=\frac{|P_{+}^{\xi }\left( k\right) |^{2}-|P_{-}^{\xi }\left(
k\right) |^{2}}{|P_{+}^{\xi }\left( k\right) |^{2}+|P_{-}^{\xi }\left(
k\right) |^{2}}.
\end{equation}%
This quantity is the difference between the absorption of the left- and
right-handed lights ($\pm $), normalized by the total absorption, around the
K$_{\xi }$ point. We show the optical polarization for the topological and
band insulators in Fig.\ref{FigPorali}. We find that all optical
polarizations are perfectly polarized at the K and K' points ($k=0$).
Namely, the selection rule holds exactly at the K and K' points, where $\eta
=\pm 1$. Then, $|\eta ^{\xi }(k)|$ decreases to $0$ as $k$ increases. It is
to be emphasized that the optical polarization of the fundamental interband
transition is opposite whether the system is a topological insulator or a
band insulator.

By neglecting the Rashba term ($\lambda _{\text{R2}}=0$), we obtain an
analytic formula of the optical polarization for the transitions $\omega
_{1} $ and $\omega _{2}$, 
\begin{equation}
\eta ^{\xi }(k)=\frac{2\xi \Delta _{s}\sqrt{4v_{\text{F}}^{2}k^{2}+\Delta
_{s}^{2}}}{4v_{\text{F}}^{2}k^{2}+2\Delta _{s}^{2}}.
\end{equation}%
The optical polarization changes the sign at the topological phase
transition $E_{z}=E_{\text{cr}}$, since $\Delta _{s}$\ changes the sign.
This property holds as it is even for $\lambda _{\text{R2}}\neq 0$, as seen
in Fig.\ref{FigPorali}.

In this paper we have analyzed optical absorption in silicene. We have shown
that silicene exhibits a strong circular dichroism obeying the spin-valley
selection rule (Fig.\ref{FigOptLevel}). The response is opposite whether
silicene is a topological or band insulator. Indeed, in the topological
(band) insulator phase, the optical field with the right-handed circular
polarization excites only up-spin (down-spin) electrons in the K valley,
while the one with the left-handed circular polarization excites only
down-spin (up-spin) electrons in the K' valley, as far as the fundamental
transition $\omega _{1}$ concerns. Now, we are able to generate a
longitudinal charge current with a definite spin by applying an in-plane
electric field. By measuring the spin direction we can tell if silicene is a
topological or band insulator.

\label{SecConclusion}

I am very much grateful to N. Nagaosa for many helpful discussions on the
subject. This work was supported in part by Grants-in-Aid for Scientific
Research from the Ministry of Education, Science, Sports and Culture No.
22740196.

\end{document}